# A low-cost cryogenic temperature measurement system using Arduino microcontroller

Woong Sung Lee

*CheongShim International Academy, Gyeonggido, Republic of Korea*

Abstract

We developed a simple, flexible, low-cost, and computer-controlled cryogenic temperature measurement system for undergraduate instructional laboratories. An Arduino microcontroller board measures the voltage across a silicon diode to calculate its temperature. Resistors and a voltage regulator provide constant current into the silicon diode. We present a graphical user interface based on the open-source Processing language. The cost of the complete temperature measurement system is thus only a small fraction of any highly-developed commercial system. Our performance test shows that the system works at a reasonable accuracy from 297.15 K (typical room temperature) down to 77 K (liquid nitrogen temperature).

# I. INTRODUCTION

Cryogenics and superconductivity education is an essential part of a typical undergraduate physics curriculum [1]. Nevertheless, the implementation of the required temperature measurement system in an instructional laboratory is typically an expensive procedure. A typical commercial temperature monitor (for example, Lake Shore Cryotronics Model 224) precisely measures temperature down to 300 mK and is ideal for multi-sensor lab uses [2]. However, the cost for such a unit could be $3000 or more, and such precise measurement is unnecessary for undergraduate educational purposes. Furthermore, complex operating procedures and bulky hardware make repair and maintenance expensive. The objective of this work is to develop a simple, low-cost cryogenic temperature measurement system to replace highly-developed commercial solutions and aid in undergraduate-level education.

This project is possible due to the Arduino microcontroller [3]. The 20 USD to 30 USD open-source Arduino hardware is a versatile yet easy-to-learn microcontroller that can run scientific instruments [4]. The Arduino microcontroller has proved ideal for educational demonstrations in previous literature by assisting in the creation of mirror mount system for optics setups [5], gamma-ray spectroscopy [6], and galvanometer system [7]. In this paper, the Arduino microcontroller was used to supply voltage into constant current source circuit and to measure voltage across the attached silicon diode. Due to straightforward operation and low cost, our cryogenic temperature measurement system can be used in setups where its implementation has been too costly and burdensome.

Along with Arduino microcontroller, a silicon diode was used as the temperature sensor due to the strong temperature dependence of its forward bias voltage drop [8]. The use of silicon diodes as temperature sensors was reported as early as the 1960s [9-10]. Silicon diodes have low cost [11] and simple voltage temperature relationship over a wide range (4.2 to 880K) [12], making them ideal for educational demonstrations. Additionally, silicon diodes can be easily integrated on chip electronics like Mr. SQUID [13], a popular superconducting quantum interference device designed for undergraduate physics lab courses.

This paper is organized as follows. In Sec. II, the basic principles of the Arduino cryogenic temperature measurement system are discussed. The electronics and software setup of the system is described in Sec. III. Experimental results obtained with the low-cost setup are presented in Sec. IV, along with a comparison of these data with a highly-developed commercial device. Section. V gives a few concluding remarks about the low-cost cryogenic temperature measurement system.

## II. BASIC PRINCIPLES

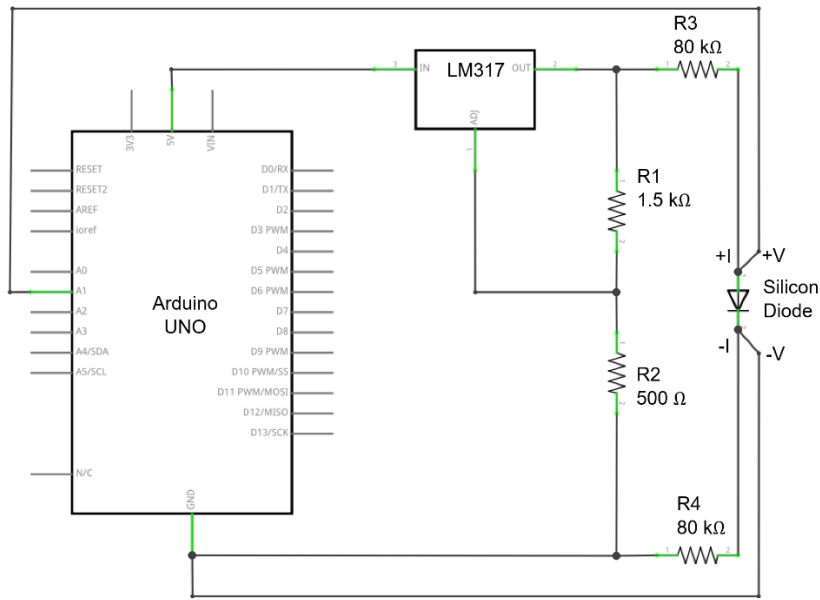

**Figure 1.** Our setup includes a constant current source of ~10 μA created with a voltage regulator and several resistors. It also includes the voltage measurement for the silicon diode.

Silicon diodes are used for temperature measurement in two different modes: (a) constant current mode and (b) constant voltage mode. Our cryogenic temperature measurement setup uses constant current mode due to its simplicity in implementation. When the diode is operated at a constant forward current, the voltage across the diode varies approximately linearly with temperature.

The voltage $V$ across the silicon diode can be expressed as

$$V = V_g - \frac{k_B T}{q}[\ln(K) + r\ln(T) - \ln(I)] \tag{1}$$

where $V_g$ is the extrapolated energy gap at 0 K, $k_B$ is the Boltzmann's constant, $T$ is the absolute temperature, $q$ is the charge of an electron, $I$ is the current across the diode, $K$ and $r$ are constants independent of temperature. Constant $K$ depends on the geometric factors of the silicon diode and constant $r$ is about ~3.5 for silicon. The detailed mathematical description is out of the scope of this paper and is given in the previous literature [14].

Eq. (1) shows that the silicon diode will display voltage approximately linear with temperature. It is important to note that at temperatures below 30 K, the diode behavior drastically changes [8]. However, most undergraduate laboratory demonstrations on cryogenics and superconductivity are done at the liquid nitrogen temperature, which is about 77 K. Thus, Eq. (1) can be expressed as

$$T = AV + B \tag{2}$$

where constants $A$ and $B$ are experimentally determined. When two calibration points are made, Eq. (2) can be rewritten as

$$T = \left(\frac{T_2 - T_1}{V_2 - V_1}\right)V + T_2 - \left(\frac{T_2 - T_1}{V_2 - V_1}\right)V_2 \tag{3}$$

Eq. (3) is the basis of our temperature measurement system. As a constant current is supplied into the silicon diode, we initially take two calibration points at known temperatures. The Arduino microcontroller records the voltage across the silicon diode at each temperature. After the microcontroller completes Eq. (3), the system can make other temperature measurements according to the computed equation of line.

### III. SETUP

In this section, we describe the setup of the electronics and the software. Even though we present the details of our components, we stress that any variation that follows our working principle is possible.

**A. Electronics**

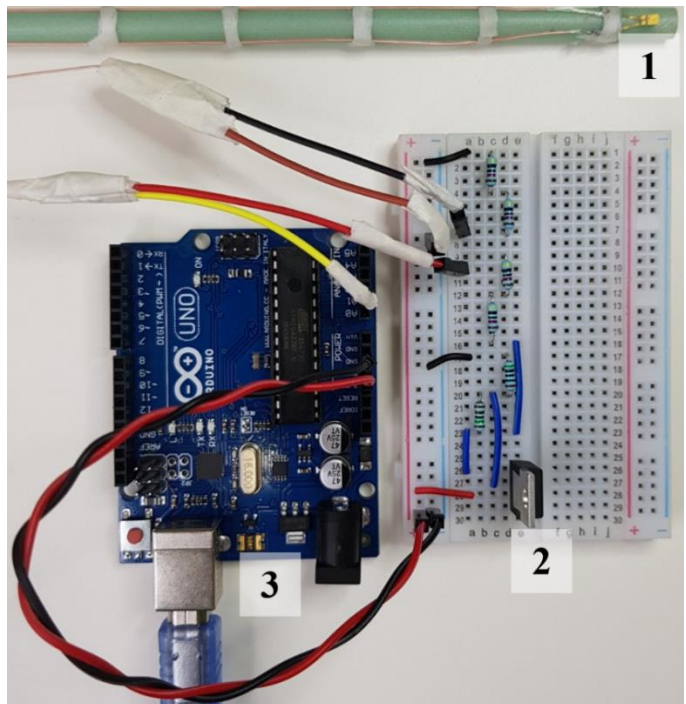

**Figure 2.** The completed electronics setup of the temperature measurement setup. The silicon diode (1) is soldered to the constant current circuit and the voltage measurement circuit. The voltage regulator (2) is a part of the constant current circuit, which is connected to the Arduino microcontroller (3).

Figure 2 shows the complete electronics setup of our temperature measurement system. In this figure, the voltage regulator (LM 317) is from Texas Instruments [15]. The voltage regulator currently costs around 2 USD. The Arduino 5 V output pin is connected to the voltage regulator to provide constant 10 µA current into the connected silicon diode. The resistors are configured as shown in figure 1 except the two 80 kΩ resistors, which were replaced with four 40 kΩ resistors. The silicon diode (DT-670) used in this setup is from Lake Shore Cryotronics [16]. However, the component can be replaced with any glass-encapsulated silicon diode. A silicon diode (1N914) from ON Semiconductor [17], currently priced under 1 USD, is a good substitute. In figure 2, the silicon diode is mounted on a 60 cm plastic rod for convenient temperature measurement in dewars.

The Arduino microcontroller runs a program available in the Appendices. The program can be uploaded to the microcontroller using Arduino Integrated Development Environment (IDE), which can be found online [3]. The Arduino IDE is available for many different computer operating systems. The Arduino code expects input from the serial port to determine its operation. This requires the USB port on the Arduino microcontroller to act as a serial port. The Arduino program assumes that the two calibration temperatures, which are explained in Sec. II, are 77 K (liquid nitrogen temperature) and 297.15 K (typical room temperature). These values can be altered to suit the user's situation.

**B. Software**

Any program with the serial interface can talk to the Arduino microcontroller. We use Processing 3.5.3 [18], which is available for most computer operating systems. Processing talks to the Arduino microcontroller through the native serial port library. In addition, the controlP5 library, which is available for most computer operating systems as well, is used to program the graphical user interface (GUI) [19]. Processing, the serial library, and the control P5 library can be installed on a computer in less than half an hour. The GUI is shown in figure 3. A flow chart illustrating its functions is shown in figure 4.

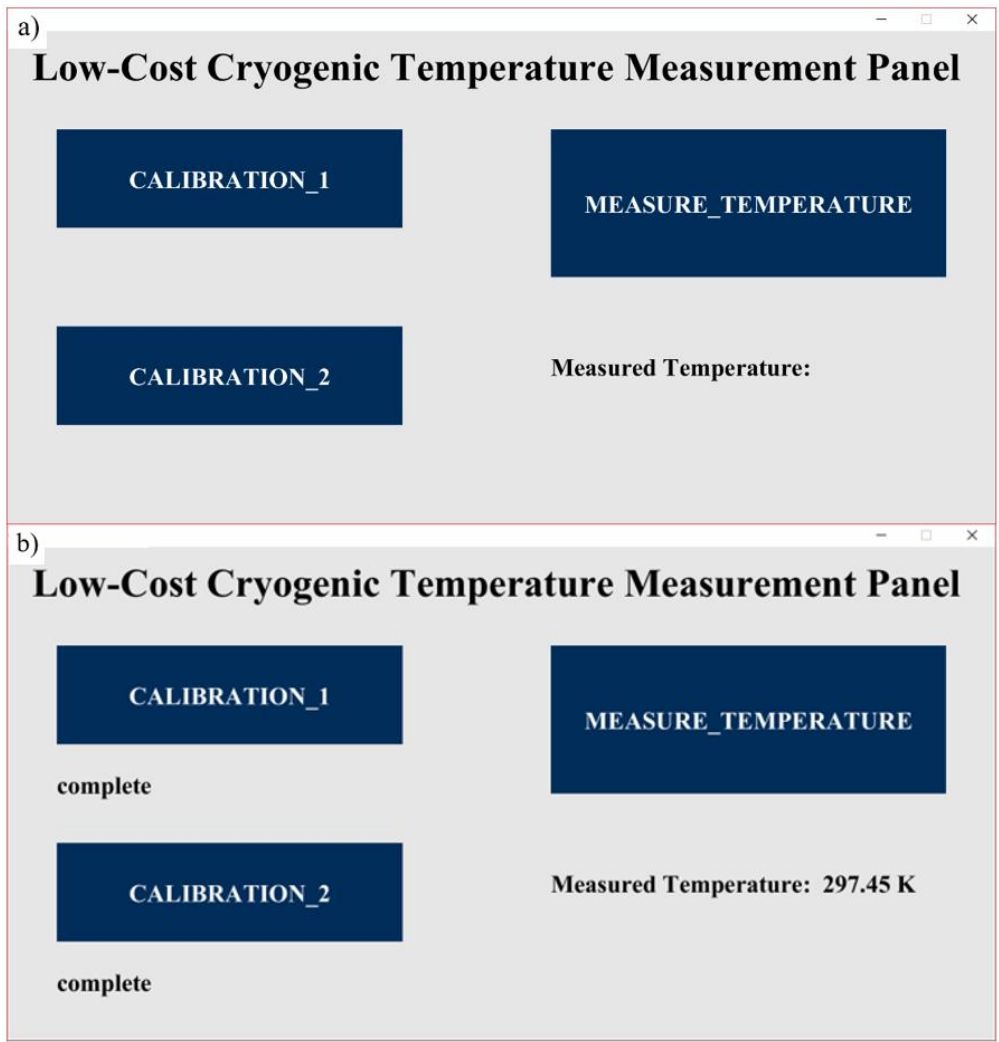

**Figure 3.** Graphical user interface implemented in Processing. (a) Before calibration. (b) After calibration.

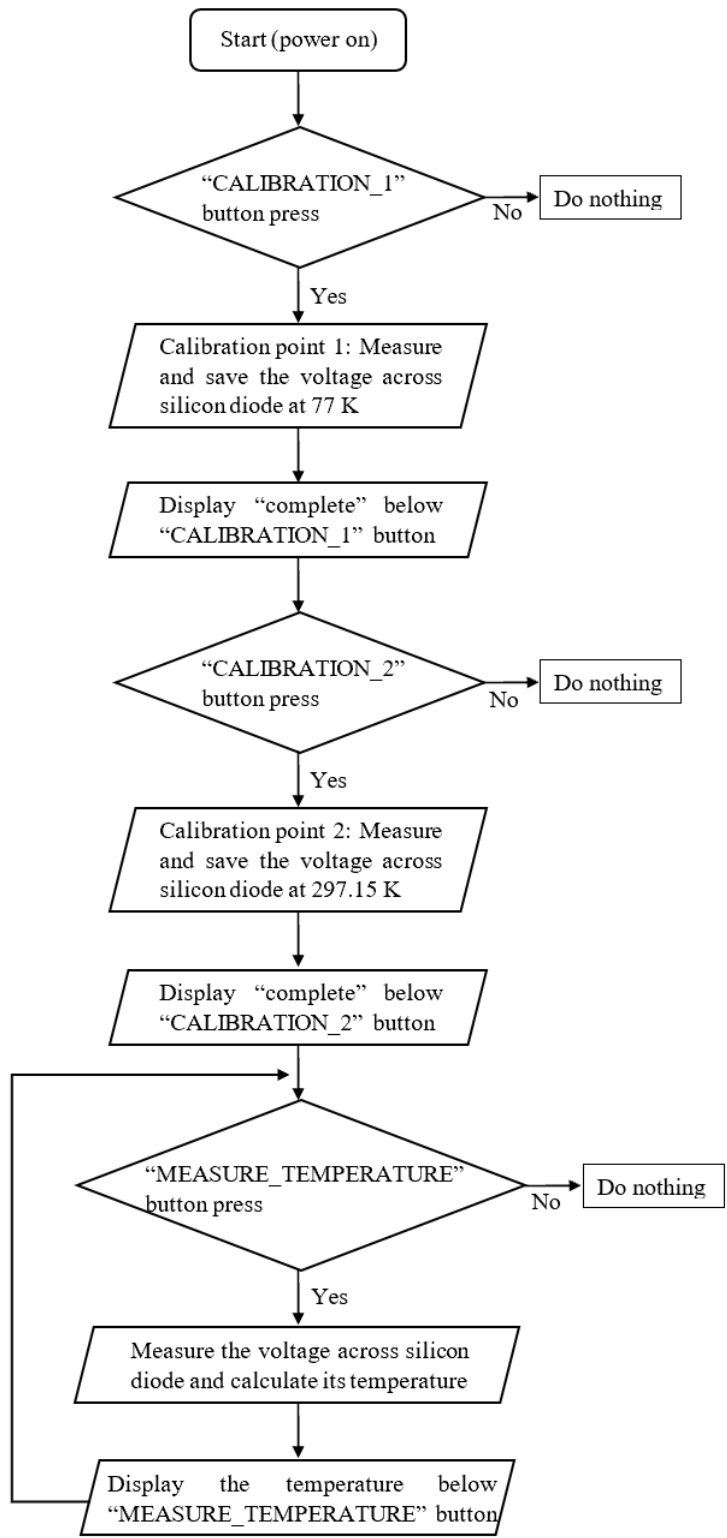

**Figure 4.** Flow chart representing the functions of the GUI

## IV. TESTING AND RESULTS

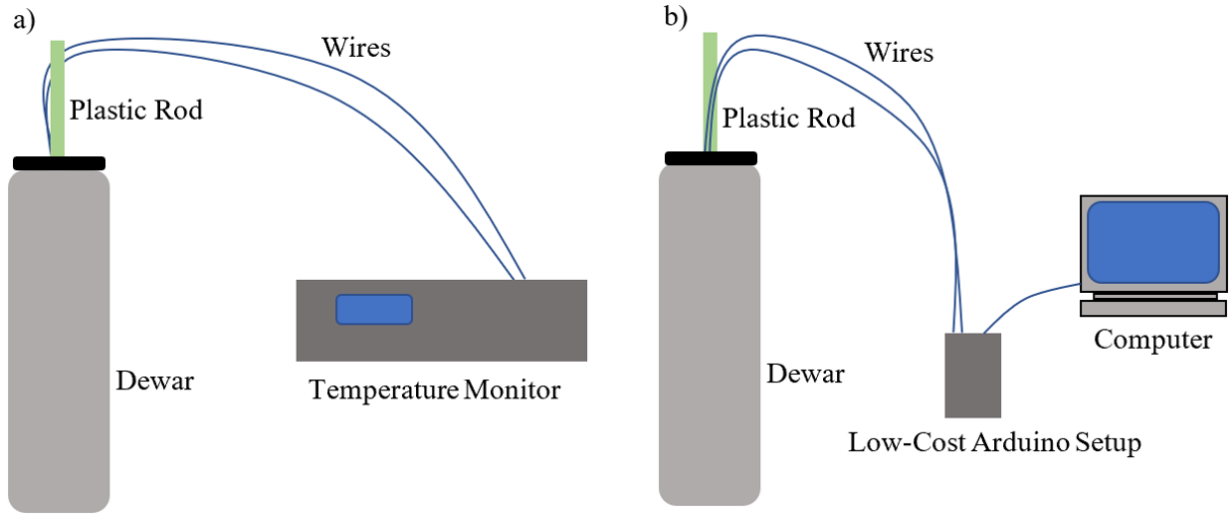

**Figure 5.** Experimental setup to test the performance of our low-cost setup. (a) The silicon diode is connected to a sensor input port of the temperature monitor. (b) The same silicon diode is connected to our low-cost Arduino setup as depicted in Figure. 2.

In order to quantitively measure the performance of our low-cost setup, we decided to compare it to a highly-developed commercial temperature controller (Model 336) from Lake Shore Cryotronics [20]. The commercial device, which currently costs around $4800, is tested to have temperature accuracy better than $\pm 100$ mK when in use with the silicon diode (DT-670). The setup is shown in figure 5. The low-cost setup is calibrated following the steps outlined in figure 4. The dewar, which is about 40 cm in height and about 10 cm in diameter, is filled with liquid nitrogen to about 15 cm. The plastic rod, which has the silicon diode mounted at the end, as shown in figure 2, is inserted into the dewar at different depths to simulate various temperatures. Starting from 5 cm, we move the plastic rod slowly into the dewar until the rod is 25 cm into the dewar, where the silicon diode is fully submerged in liquid nitrogen. Temperature measurement is made every 2.5 cm point. For accuracy, we waited for the temperature value to stabilize at each point.

The experimental results are depicted as a line graph in figure 6. Several tests confirm that, from room temperature down to liquid nitrogen temperature, the maximum difference between the measurement from the low-cost setup and the measurement from the commercial temperature controller is about 2 K.

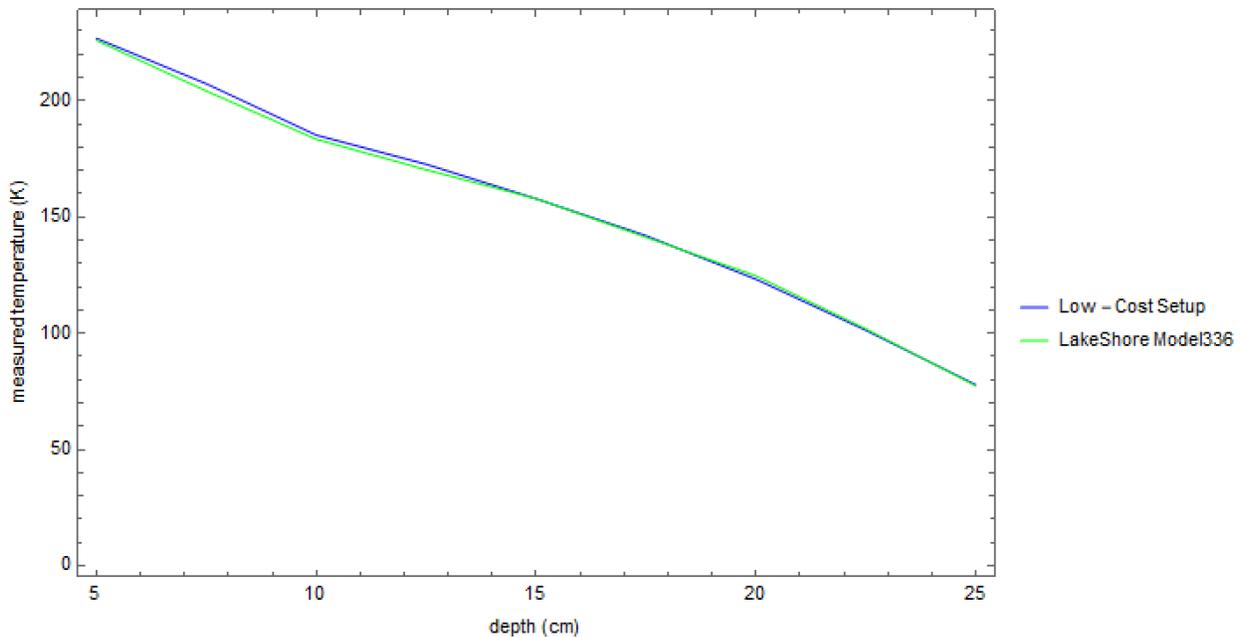

**Figure 6.** Measured values from our low-cost setup and the commercial device. The points are connected in lines for readability.

## V. CONCLUSION

All features of the low-cost cryogenic temperature measurement system are working as expected, and it can be used for educational demonstrations in undergraduate instructional laboratories. By soldering the components that are connected by the solderless breadboard in figure 2, connections between the components can be improved for higher accuracy in temperature measurements. In addition, any exposed wire can be covered with insulation tape to prevent external interruption to the voltage reading of the Arduino microcontroller. In Sec. IV, performance test showed that our low-cost setup operates with reasonable accuracy from the typical room temperature down to the liquid nitrogen temperature. Such a measurement range is adequate for undergraduate cryogenics and superconductivity education, which involves high-temperature superconductors. In Sec. II, it is theoretically predicted that our low-cost setup operates down to 30 K, if needed. For temperature under 30 K, completely different program design is needed.

**ACKNOWLEDGEMENTS**

W. S. Lee wishes to thank Dr. Andrei Matlashov for his valuable support in the laboratory and with the manuscript. Associate Professor Carlotta Berry, Dr. Hyeon-soo Ryu, Jaesung Yoon, Geunjung Lee, and Eunsoo Shin are gratefully acknowledged for their help in the manuscript. This work was supported by the Korea Undergraduate/graduate/high-school Science Program (KUSP) at Center for Axion and Precision Physics Research (CAPP), Institute of Basic Science, Republic of Korea.

**REFERENCES**

[1] Universities offering Cryogenics and Superconductivity education https://cryogenicsociety.org/cryo_careers/universities_offering_cryogenics_and_superconductivity_education_in_the_united_states/

[2] Lake Shore Model 224 product overview web page http://www.lakeshore.com/products/categories/overview/temperature-products/cryogenic-temperature-monitors/model-224-temperature-monitor

[3] Arduino web site http://www.arduino.cc

[4] J. M. Pearce 2012 "Building research equipment with free, open-source hardware," *Science* **337(6100)** 1303

[5] Maithreyi Gopalakrishnan and Markus Gühr 2015 A low-cost mirror mount control system for optics setups *Am. J. Phys.* **83** 186

[6] C. M. Lavelle 2018 Gamma ray spectroscopy with Arduino UNO *Am. J. Phys.* **86** 384

[7] Jen-Feng Hsu, Shonali Dhingra, and Brian D'Urso 2017 Design and construction of a cost-efficient Arduino-based mirror galvanometer system for scanning optical microscopy *Am. J. Phys.* **85** 68

[8] Mohtashim Mansoor, Ibraheem Haneef, Suhail Akhtar, Andrea De Luca, and Florin Udrea 2015 Silicon diode temperature sensors – A review of applications *Sens. and Actuators A: Phys.* **232** 63

[9] H. Harris 1961 Concerning a thermometer made with solid-state diodes *Sci. Am.* **204** 192

[10] A. G. McNamara 1962 Semiconductor diodes and transistors as electrical thermometers *Rev. Sci. Instrum.* **33** 330

[11] P.R.N. Childs, J.R. Greenwood, and C.A. Long 2000 Review of temperature measurement *Rev. Sci. Instrum.* **71** 2959

[12] N.S. Boltovets *et al.* 2001 Ge-film resistance and Si-based diode temperature microsensors for cryogenic applications *Sens. Actuators, A: Phys*. **92** 191

[13] Mr. SQUID Brochure https://starcryo.com/mr-squid/

[14] Florin Udrea, Sumita Santra, and Julian W. Gardner 2008 CMOS temperature sensors – concepts, state-of-the-art and prospects *2008 International Semiconductor Conference* 31

[15] LM317 3-Terminal Adjustable Regulator, Texas Instruments http://www.ti.com/lit/ds/slvs044x/slvs044x.pdf

[16] DT-670 silicon diode, Lake Shore Cryotronics http://www.lakeshore.com/products/categories/overview/temperature-products/cryogenic-temperature-sensors/dt-670-silicon-diodes

[17] 1N914 small signal diode, ON Semiconductor http://www.onsemi.com/pub/Collateral/1N914-D.PDF

[18] Processing website http://processing.org

[19] controlP5 library website http://www.sojamo.de/libraries/controlP5/

[20] Model 336 Temperature Controller, Lake Shore Cryotronics https://www.lakeshore.com/products/categories/overview/temperature-products/cryogenic-temperature-controllers/model-336-cryogenic-temperature-controller

# APPENDIX A – Arduino Code (for future supplementary material)

```
//Measure voltage across the silicon diode
int Vpin = A1;

//Declare required variables
float voltage;
float V1;
float V2;
float V3;
float T;

//Recommended calibration temperatures in Kelvin scale
//These temperatures can change to the user's preference
float T1 = 77;
float T2 = 297.15;

void setup() {
//Load serial monitor
 Serial.begin (9600);
}

void loop() {
 if(Serial.available()){
   char val = Serial.read();

//Save the first calibration voltage when CALIBRATION_1 button is pressed
   while(val == 'a'){
     voltage=analogRead(Vpin);
     V1 = voltage/1023*5.00;
     val = 'd';
   }

//Save the second calibration voltage when CALIBRATION_2 button is pressed
   while(val == 'b'){
     voltage=analogRead(Vpin);
     V2 = voltage/1023*5.00;
     val = 'd';
   }
//Calculate temperature when MEASURE_TEMPERTURE button is pressed
   while(val == 'c'){
     voltage=analogRead(Vpin);
     V3 = voltage/1023*5.00;
    T = ((T2-T1)/(V2-V1))*V3 + T2 - ((T2-T1)/(V2-V1))*V2;
     Serial.println(T);
     val = 'd';
   }
 }
}
```

# APPENDIX B – Processing Code (for future supplementary material)

```
//Setting up
import controlP5.*;
import processing.serial.*;
String m_="";
String m_2="";
String c_1="";
String c_2="";
Serial port;

//create cp5 object
ControlP5 cp5;
PFont font;
PFont font2;

void setup (){
//Setting the window size and ports
  size(2000,1000);
  printArray(Serial.list());
  port = new Serial(this, "COM13", 9600);
  port.bufferUntil('\n');

//Setting "CALIBRATION_1", "CALIBRATION_2", "MEASURE_TEMPERATURE" buttons
  cp5 = new ControlP5(this);
  font = createFont("times new roman bold", 50);
  font2 = createFont("times new roman bold", 80);
  cp5.addButton("calibration_1")
    .setPosition(100,200)
    .setSize(700,200)
    .setFont(font)
  ;
  cp5.addButton("calibration_2")
    .setPosition(100,600)
    .setSize(700,200)
    .setFont(font)
  ;
  cp5.addButton("measure_temperature")
    .setPosition(1100,200)
    .setSize(800,300)
    .setFont(font)
  ;
}

void draw (){
//Setting the window's background color and label
  background(230,230,230);
  fill(0,0,0);
  textFont(font2);
  text("Low-Cost Cryogenic Temperature Measurement Panel",50,100);
  textFont(font);
```

```
 //Setting location to show "complete" message and measured temperature
  text(c_1, 100, 500);
  text(c_2, 100, 900);
  text("Measured Temperature:", 1100, 700);
  text(m_2, 1650, 700);
 }

//Defining "CALIBRATION_1" function
void calibration_1(){
 port.write("a");
 c_1 = "complete";

}

//Defining "CALIBRATION_2" function
void calibration_2(){
 port.write("b");
 c_2 = "complete";
  }

//Defining function "MEASURE_TEMPERATURE"
void measure_temperature(){
 m_="";
 m_2="";
 port.write("c");
 while(port.available()>0){
  m_ = port.readStringUntil('\n');
  m_2 =  m_ + "K";
  }
}
```